\documentclass[reprint,twocolumn, tightenlines]{revtex4-1}

\usepackage{amsthm}
\usepackage{amsmath}
\usepackage{bbm, dsfont}
\usepackage{graphicx}
\usepackage[usenames,dvipsnames]{color}
\usepackage[colorlinks=true,citecolor=blue,urlcolor=black]{hyperref}
\usepackage[table]{xcolor}
\usepackage{colortbl}
\usepackage{color}
\usepackage{multirow}
\usepackage{hhline}
\usepackage{tabu}
\usepackage{epstopdf}
\usepackage{booktabs}

\newcolumntype{C}[1]{ >{ \centering\arraybackslash}p{#1}}
\def \minipagesize {4.2cm}

\newcommand{\dt}[3]{\cellcolor{#1} \scriptsize #2}

\newcommand{\tn}[1]{\textnormal{#1}}
\newcommand{\be}{\begin{equation}}
\newcommand{\ee}{\end{equation}}

\newcommand{\del}[1]{}

\newcommand{\sket}[1]{{\ensuremath{\lvert#1\rangle}}}
\newcommand{\lket}[1]{{\ensuremath{\left\lvert#1\right\rangle}}}
\newcommand{\ket}[1]{\if@display\lket{#1}\else\sket{#1}\fi}

\newcommand{\sbra}[1]{{\ensuremath{\langle#1\rvert}}}
\newcommand{\lbra}[1]{{\ensuremath{\left\langle#1\right\rvert}}}
\newcommand{\bra}[1]{\if@display\lbra{#1}\else\sbra{#1}\fi}

\newcommand{\sbraket}[2]{{\ensuremath{\langle#1\rvert#2\rangle}}}
\newcommand{\lbraket}[2]{{\ensuremath{\left\langle#1\!\left\rvert\vphantom{#1}#2\right.\!\right\rangle}}}
\newcommand{\braket}[2]{\if@display\lbraket{#1}{#2}\else\sbraket{#1}{#2}\fi}

\newcommand{\sketbra}[2]{{\ensuremath{\lvert #1\rangle\!\langle #2\rvert}}}
\newcommand{\lketbra}[2]{{\ensuremath{\left\lvert #1\right\rangle\!\!\left\langle #2\right\rvert}}}
\newcommand{\ketbra}[2]{\if@display\lketbra{#1}{#2}\else\sketbra{#1}{#2}\fi}

\newcommand{\proj}[1]{\ketbra{#1}{#1}}

\definecolor{gray4}{gray}{0.8}
\definecolor{gray2}{gray}{0.6}

\begin{document}
\title{Detector-device-independent quantum key distribution}
\author{Charles Ci Wen \surname{Lim}}\email{ciwen.lim@unige.ch}
\author{Boris Korzh}\email{boris.korzh@unige.ch}
\author{Anthony Martin}
\author{F\'elix Bussi\`eres}
\author{Rob Thew} 
\author{Hugo Zbinden}
\affiliation{Group of Applied Physics, University of Geneva, Chemin de Pinchat 22, CH-1211 Geneva 4, Switzerland}

\begin{abstract}


Recently, a quantum key distribution (QKD) scheme based on entanglement swapping, called measurement-device-independent QKD (mdiQKD), was proposed to bypass all detector side-channel attacks. While mdiQKD is conceptually elegant and offers a supreme level of security, the experimental complexity is challenging for practical systems. For instance, it requires interference between two widely separated independent single-photon sources, and the rates are dependent on detecting two photons - one from each source. Here we experimentally demonstrate a QKD scheme that removes the need for a two-photon system and instead uses the idea of a two-qubit single-photon (TQSP) to significantly simplify the implementation and improve the efficiency of mdiQKD in several aspects. 

\end{abstract}

\maketitle


Quantum key distribution (QKD) enables the exchange of cryptographic keys between two separated users, Alice and Bob, who are connected by a potentially insecure quantum channel~\cite{bb84, Gisin2002, Scarani2009, Lo2014}. Unlike conventional key distribution schemes, the security of QKD depends only on the principles of quantum physics and can be proven information-theoretically secure.~However, despite the potential of QKD, one still has to be prudent about potential side-channel attacks that may lead to security failures.~For example, it has been shown that with detector blinding techniques, it is possible to remotely hack the measurement unit of some QKD systems~\cite{Lydersen2010}. Although it is possible to implement appropriate countermeasures for specific attacks, one may be wary that the adversary, Eve, could devise new detector control strategies, unforeseen by the users.

To prevent all known and yet-to-be-discovered detector side-channel attacks, a measurement-device-independent~QKD (mdiQKD) protocol was proposed~\cite{Lo2012}. In this scheme, Alice and Bob each randomly prepare one of the four Bennett Brassard (BB84) states~\cite{bb84} and send it to a third party, Charlie, whose role is to introduce entanglement between Alice and Bob via a Bell-state measurement (BSM)~\cite{Biham1996, Inamori2008}. Obviously, Alice and Bob do not have to trust Charlie since any other non-entangling measurement would necessarily introduce some noise between them. In practice, mdiQKD can be implemented with phase-randomized weak coherent (BB84) states (WCSs), using either time-bin encoded qubits \cite{Rubenok2013, Liu2013} or polarization-encoded qubits \cite{Silva2013, Tang2014}. To meet the assumption that Alice and Bob send single photons, as required by mdiQKD, they randomly vary the intensity of their laser pulses and use the decoy-state method~\cite{Hwang2003,Lo2005b, Wang2005} to estimate the fraction of single-photon states sent to and detected by Charlie.

Unfortunately, mdiQKD possesses many drawbacks. Firstly, the achievable secure key rates (SKR) are significantly lower compared to conventional prepare and measure (P\&M) QKD systems~\cite{Dixon2010, Tanaka2012, Walenta2014, Korzh2014c}. This is mainly because a two-photon BSM relies on coincidence detections, meaning that the SKR scales with $(\eta_{det}P_{1}(\mu) )^2$, where $\eta_{det}$ is the single photon detector (SPD) efficiency and $P_{1}(\mu)$ is the probability of the source emitting a single-photon~\footnote{For a WCS $P_{1}(\mu)=\mu e^{-\mu}$, where $\mu < 1$ is the average photon number per pulse. Typical values of $\eta_{det}$ are $10-30\%$ for practical InGaAs SPDs.}. Another factor is that a two-photon BSM implemented with linear optics is at most 50\% efficient~\cite{Vaidman1999, Lutkenhaus1999, Valivarthi2014} and, when using WCSs, the results from one of the bases cannot be used for the raw-key generation due to an inherent 25\% error rate~\cite{Rubenok2013, Silva2013}. Furthermore, the resource overhead in the finite-key scenario~\cite{Curty2014} is significantly larger compared to common P\&M schemes~\cite{Lim2014, Korzh2014c}, due to the need to apply the decoy-state method twice (once for each source), increasing the statistical fluctuations. For example, at 150~km, Alice and Bob would have to send at least $10^{14}$ laser pulses to Charlie before key distillation is possible~\cite{Curty2014}. Finally, the technological complexity of mdiQKD is greater due to the use of two-photon interference, requiring both photons to be indistinguishable in all degrees of freedom (DOFs): temporal, polarization and frequency.  

Here we report on the implementation of a QKD scheme that overcomes the aforementioned limitations but is still secure against all detector side-channel attacks. This bridges the gap between the superior performance and practicality of P\&M QKD schemes and the enhanced security offered by mdiQKD. Note that a similar scheme, following the same basic idea, has been proposed elsewhere~\cite{Gonzalez2014}. Our scheme, henceforth referred to as detector-device-independent QKD (ddiQKD), essentially follows the idea of mdiQKD, however, instead of encoding separate qubits into two independent photons, we exploit the concept of a two-qubit single-photon (TQSP). This scheme has the following advantages: (1) it requires only single-photon interference, (2) the linear-optical BSM is 100\% efficient~\cite{Boschi1998}, (3) the secret key rate scales linearly with the SPD detection efficiency and (4) it is expected that in the finite-key scenario the minimum classical post-processing size is similar to that of P\&M QKD schemes. In the following we outline the main concepts and demostrate a proof-of-principle experiment.
\begin{figure}[t!] 
\includegraphics[width=85mm]{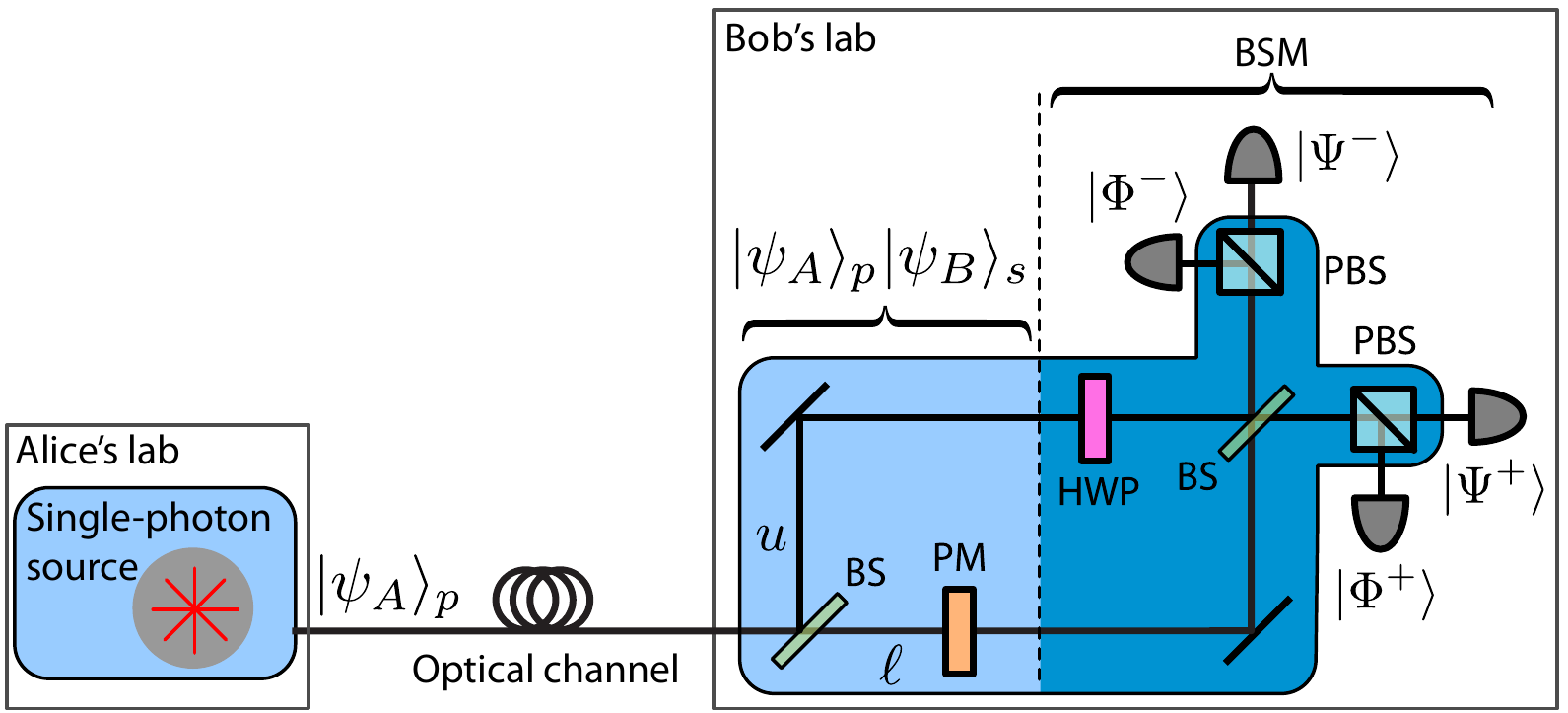}
\caption{The conceptual setup. Alice encodes her qubit $\ket{\psi_A}_p$ in the polarization DOF of a single photon, sends it to Bob who encodes his qubit $\ket{\psi_B}_s$ in the spatial DOF using a 50/50 beam splitter (BS) and a phase modulator (PM). Bob then performs a complete and deterministic Bell-State measurement (BSM) on both qubits using a half-wave plate (HWP), polarizing beam splitters (PBS) and single-photon detectors (SPDs). Components inside the shaded regions of Alice and Bob's labs are trusted devices, whilst the SPDs are untrusted. 
}\label{fig:concept}
\end{figure}

The protocol works as follows; see Fig.~\ref{fig:concept}. Alice first prepares a single photon in the qubit state $\ket{\psi_A}_p$ chosen at random from the following set of BB84 states: 
\begin{equation*}
\ket{\psi_A}_p\in_r\left\{ 
  \begin{array}{l }
    \ket{+}\phantom{i}=\frac{1}{\sqrt{2}}(\ket{H}+\ket{V}), \\
    \ket{-}\phantom{i}=\frac{1}{\sqrt{2}}(\ket{H}-\ket{V}), \\
    \ket{+i}=\frac{1}{\sqrt{2}}(\ket{H}+i\ket{V}), \\
    \ket{-i}=\frac{1}{\sqrt{2}}(\ket{H}-i\ket{V}),
  \end{array} \right.
\end{equation*} 
where the subscript $p$ indicates this is a qubit in the polarization DOF of the photon. Alice sends $\ket{\psi_A}_p$ to Bob via an untrusted quantum channel. Upon reception of the photon, Bob encodes his random qubit state $\ket{\psi_B}_s$ in the spatial DOF (hence the subscript ``$s$''). To achieve this, Bob sends the photon to a 50/50 beam splitter (BS). We denote $\ket{u}$ and $\ket{\ell}$ the states of the basis defined by the ``upper'' and ``lower''arms after the BS, respectively. He then applies a phase $\varphi$ chosen at random in the set $\{0,\pi/2,\pi,3\pi/2\}$ on the lower arm to prepare the state $\ket{\psi_B}_s = (\ket{u} + \mathrm{e}^{i\varphi}\ket{\ell})$, yielding BB84 states in the spatial modes. Both DOFs have so far been created and manipulated independently of each other, and thus the two-qubit state can be written as $\ket{\psi_A}_p\otimes\ket{\psi_B}_s$. 
  
We then define the following Bell states:
\begin{eqnarray}
\ket{\Phi^{\pm}} &=& \frac{1}{\sqrt{2}}(\ket{H}_p\ket{u}_s\pm \ket{V}_p\ket{\ell}_s), \label{bell1}\\
\ket{\Psi^{\pm}} &=& \frac{1}{\sqrt{2}}(\ket{H}_p\ket{\ell}_s\pm \ket{V}_p\ket{u}_s). \label{bell2}
\end{eqnarray}
A complete and deterministic BSM of these states is realized by first applying the unitary transformation $\ket{Hu}\rightarrow \ket{Vu}$ and $\ket{Vu}\rightarrow \ket{Hu}$ on the upper arm using a half-wave plate (HWP), followed by recombination of the arms on a 50/50 BS, and finally by a projection in the $\{\ket{H},\ket{V}\}$ basis using two PBSs on the two output arms followed by four SPDs. In this way, a click on each SPD corresponds to a projection on one of the four Bell states; see Fig.~\ref{fig:concept}. 

To show how the raw key establishment functions, let us first define the mutually unbiased bases $\mathcal{B}_{X} = \{\ket{+},\ket{-}\}$ and $\mathcal{B}_{Y} = \{\ket{+i},\ket{-i}\}$. The bit to be established is encoded in Alice's state, i.e.~$\ket{+}$ and $\ket{+i}$ encode bit~0, and $\ket{-}$ and $\ket{-i}$ encode bit~1. After the measurement phase, Bob uses an authenticated channel to announce the success of the BSM and reveals the basis he used to encode his qubit. Subsequently, Alice announces whether Bob's basis choice was compatible with hers. Bob can then determine Alice's bit value according to Table~\ref{tab:correlation}, which shows all of the possible combinations.~For example, if $\ket{\psi_B}_s = \ket{+}$, the bit is 0 if he detected $\ket{\Phi^+}$ or $\ket{\Psi^+}$, and 1 otherwise. If more than one detector clicked, Bob announces a successful BSM and assigns a random bit value. Importantly, the knowledge of the bases used by Alice and Bob, along with which of the Bell states Bob obtained, does not reveal Alice's bit. Hence, Eve does not gain information on the key by controlling Bob's detectors. 
\begin{table}[t!] 
\begin{tabular}{*{4}{c}}
\begin{minipage}{\minipagesize}
\begin{tabular}{C{0.4cm}*{2}{C{0.9cm}}!{\color{black}\vrule}*{2}{C{0.9cm}}}
 \arrayrulecolor{white}
 a) & \multicolumn{4}{c}{$\ket{\Phi^+}$} \\ 
 & $+$ &  $-$ & $+i$ & $-i$\\
$+\phantom{i}$ &\dt{gray2}{0.49}{2}&\dt{white}{0.01}{1}&  \dt{gray4}{0.25}{1}  &  \dt{gray4}{0.26}{1}\\ \hline 
 \arrayrulecolor{black}
$-\phantom{i}$ & \dt{white}{0.01}{0} &\dt{gray2}{0.50}{2} &  \dt{gray4}{0.25}{1} &  \dt{gray4}{0.27}{1}\\ \hline 
 \arrayrulecolor{white}
$+i$ & \dt{gray4}{0.27}{2} &  \dt{gray4}{0.26}{2} & \dt{White}{0.01}{0} & \dt{gray2}{0.48}{2}  \\ \hline
$-i$ & \dt{gray4}{0.24}{1} &  \dt{gray4}{0.23}{1} & \dt{gray2}{0.50}{2} & \dt{white}{0.01}{0}  \\ \hline
\end{tabular}
\end{minipage} &
\begin{minipage}{\minipagesize}
\begin{tabular}{C{0.4cm}*{2}{C{0.9cm}}!{\color{black}\vrule}*{2}{C{0.9cm}}}
 \arrayrulecolor{white}
 b) & \multicolumn{4}{c}{$\ket{\Psi^+}$} \\ 
 & $+$ &  $-$ & $+i$ & $-i$\\
$+\phantom{i}$& \dt{gray2}{0.49}{3} & \dt{white}{0.02}{1} &  \dt{gray4}{0.25}{2} &  \dt{gray4}{0.27}{2}\\ \hline
 \arrayrulecolor{black}
$-\phantom{i}$ & \dt{white}{0.00}{0} &\dt{gray2}{0.50}{2} &  \dt{gray4}{0.27}{1} &  \dt{gray4}{0.24}{1} \\ \hline
 \arrayrulecolor{white}
$+i$ & \dt{gray4}{0.29}{2} &  \dt{gray4}{0.23}{1} & \dt{gray2}{0.49}{2} & \dt{white}{0.00}{0}   \\ \hline
$-i$ & \dt{gray4}{0.23}{2} &  \dt{gray4}{0.25}{2} & \dt{white}{0.01}{0} & \dt{gray2}{0.55}{2}  \\ \hline
\end{tabular}
\end{minipage}\\
&\\
\begin{minipage}{\minipagesize}
\begin{tabular}{C{0.4cm}*{2}{C{0.9cm}}!{\color{black}\vrule}*{2}{C{0.9cm}}}
 \arrayrulecolor{white}
 c) & \multicolumn{4}{c}{$\ket{\Psi^-}$} \\ 
 & $+$ &  $-$ & $+i$ & $-i$\\
$+\phantom{i}$& \dt{white}{0.00}{0} & \dt{gray2}{0.48}{2}  &  \dt{gray4}{0.28}{2} &  \dt{gray4}{0.25}{1}\\ \hline
 \arrayrulecolor{black}
$-\phantom{i}$ &  \dt{gray2}{0.54}{2} & \dt{white}{0.00}{1} &  \dt{gray4}{0.25}{1} &  \dt{gray4}{0.23}{1} \\ \hline
 \arrayrulecolor{white}
$+i$ & \dt{gray4}{0.25}{2} &  \dt{gray4}{0.26}{2} &\dt{white}{0.01}{0} &\dt{gray2}{0.52}{2}   \\ \hline
$-i$ & \dt{gray4}{0.26}{1} &  \dt{gray4}{0.24}{1} &  \dt{gray2}{0.50}{2} & \dt{white}{0.01}{0}\\ \hline
\end{tabular}
\end{minipage} &
\begin{minipage}{\minipagesize}
\begin{tabular}{C{0.4cm}*{2}{C{0.9cm}}!{\color{black}\vrule}*{2}{C{0.9cm}}}
 \arrayrulecolor{white}
 d) & \multicolumn{4}{c}{$\ket{\Phi^-}$} \\ 
 & $+$ &  $-$ & $+i$ & $-i$\\
$+\phantom{i}$& \dt{white}{0.00}{0} & \dt{gray2}{0.47}{2}  &  \dt{gray4}{0.25}{2} &  \dt{gray4}{0.25}{2}\\ \hline
 \arrayrulecolor{black}
$-\phantom{i}$ &  \dt{gray2}{0.54}{2} & \dt{white}{0.00}{0}&  \dt{gray4}{0.23}{1} &  \dt{gray4}{0.25}{1} \\ \hline
 \arrayrulecolor{white}
$+i$ & \dt{gray4}{0.26}{1} &  \dt{gray4}{0.26}{1} & \dt{gray2}{0.48}{2} & \dt{white}{0.00}{0}  \\ \hline
$-i$ & \dt{gray4}{0.26}{1} &  \dt{gray4}{0.21}{1} & \dt{white}{0.00}{0}& \dt{gray2}{0.56}{2} \\ \hline
\end{tabular}
\end{minipage}\\
\end{tabular}
\caption{Theoretical and experimentally observed probabilities for each Bell state. Rows and columns correspond to Alice's and Bob's states $\ket{\psi_A}_p$ and $\ket{\psi_B}_s$, respectively. Given a certain Bell state $k$, for each $\ket{\psi_A}_p$ there are four possible $\ket{\psi_B}_s$: white cells should happen with probability $\Pr[k]=0$, light grey cells with $\Pr[k]=1/4$ and dark grey cells with $\Pr[k]=1/2$. The experimentally observed probabilities are written in each cell.}\label{tab:correlation}
\end{table}

From a security point of view, it is important to consider carefully the operation of Bob's device. Strictly speaking, the mathematical description of his qubit, outlined previously, holds only if the light state entering the first BS is a single-photon excitation of a single optical-temporal mode. As with any other QKD scheme, it is not possible to guarantee this. Indeed, Eve may send multi-photon states through the quantum channel and break the qubit description. However, such an attack is only detrimental if she can interact with Bob's prepared states, for instance, by making unambiguous state discrimination measurements on them~\cite{Scarani2004}. This is not possible since the adversary can only interact with Alice's qubits. Additionally, if the input is a multi-photon state, with very high probability, more than one detector clicks, in which case Bob would pick a random bit value, increasing the errors in the raw bit string. This is due to the fact that the optical linear circuit of the BSM randomizes the encoded state. 

The security of our scheme requires that the final light state (just before the SPDs), taken over all possible encoding choices, is independent of the input light state. In particular, for any input state with a given $n$-photon excitation, the average final state after passing through the linear optical circuit is a fixed state. This requirement is in fact similar to the one used in the security analysis of BB84, where the average of the BB84 states has to be independent of the basis choice~\cite{Gottesman2004}. Once this requirement is met, the security of the scheme can be obtained following proof techniques for the BB84 QKD scheme. A common method to prove the security of P\&M QKD schemes is to consider an equivalent entanglement-based version, where Alice and Bob make random measurements on bipartite quantum states distributed by the adversary. To this end, we point to a formalism that allows us to see Bob's linear optical circuit as random measurements made on some entangled bipartite state.

First, let us relate the two different DOFs, i.e. $A_p$, $B_p$ denoting the polarization states of Alice and Bob respectively, while $B_s$ denotes Bob's spatial state.~Since Alice is able to prepare the four polarization BB84 states correctly, it is equivalent to consider the entanglement based version, where Alice first prepares a two-qubit maximally entangled state, $\ket{\Phi^+}$, and then performs a projective measurement on one half of the state to prepare the other half for Bob.~Mathematically, we have, $M_x \otimes \mathbbm{I} \proj{\Phi^+}_{A_p B_p} \otimes \proj{s}_{B_s}$,
where $M_x$ is the positive-operator valued measure (POVM) element corresponding to preparation $x\in \{+,-,+i,-i\}$ and $\ket{s}_{B_s}$ is an auxiliary state related to the spatial DOF.

Second, Alice sends the quantum systems $B_p$ and $B_s$ using a single photon through the quantum channel to Bob.~At this point, the resulting state is not necessarily a single photon state; it may be a multi-photon state.~In this case, the state, after tracing out system $A_p$, is described by a bipartite density operator, $\rho_{C_p C_s}$, whose dimension is unknown but fixed, i.e. it could be any $n$-photon light state.~Note that we changed the subscript $B$ to $C$ to reflect the action of the quantum channel. To proceed, we use a result from Ref.~\cite[Lemma.~1]{Beaudry2013}, which says that if, for any input state, the linear optical circuit (parameterized by $\varphi$) outputs a state that is fixed on the average, then the encoding can be seen as a purified measurement acting on the same input state and one half of a bipartite pure state, where the other half of the bipartite is the same output state. More formally, let the linear optical circuit be described by a set of completely positive trace-preserving maps, $\{\mathcal{E}_\varphi\}_\varphi$, taking the input quantum system $C_s$ to an output quantum system $D_s$, such that for any input quantum state $\rho_{C_s}$, the output quantum state is fixed over all possible encoding choices, i.e. $1/4\sum_\varphi\mathcal{E}_\varphi(\rho_{C_s})=\rho_{D_s}$ for any $\rho_{C_s}$. Then, the linear optical circuit is equivalent to making a joint measurement $\{F_{{C_s K_s}}^\varphi\}_\varphi$ on the same input state, $\rho_{C_s}$, and one half of a bipartite pure state, $\ket{\sigma}_{K_s D_s}$, living in a joint quantum system $K_s \otimes D_s$,  where the other half gives the fixed state $\rho_{D_s}$. Therefore, the purification provides a method to analyze the security of our scheme in an entanglement-based picture,  where Alice makes random BB84 measurements on one half of a bipartite quantum state, and Bob makes random purified measurements on the other half. 
\begin{figure}[t!]
\includegraphics[width=85mm]{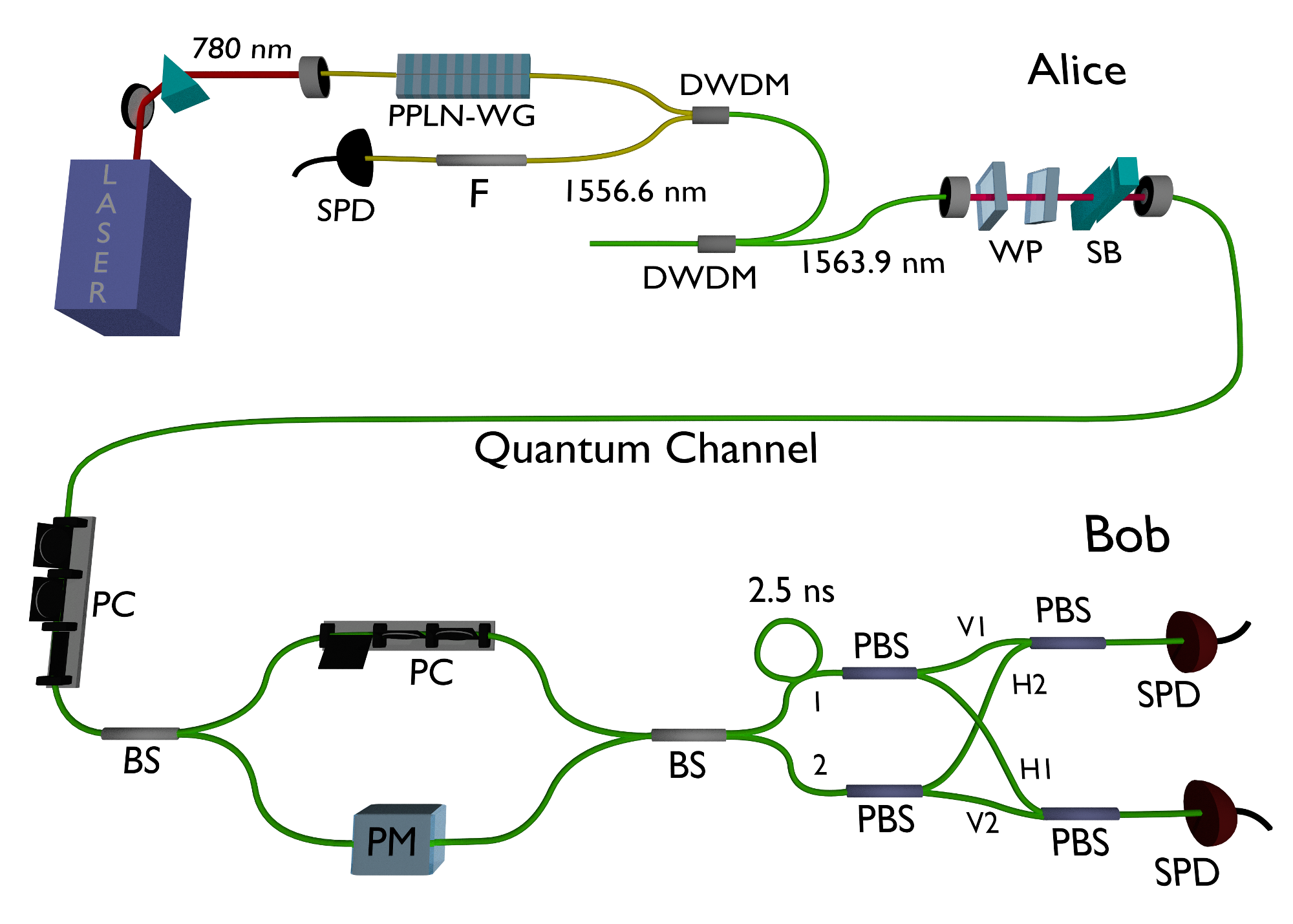}
\caption{Experimental realization of the ddiQKD protocol. Labelled components include, dense wavelength division multiplexers (DWDM), bandpass filter (F), waveplates (WP), Soleil-Babinet compensator (SB), polarization controllers (PC), phase modulator (PM), 50/50 beam splitters (BS), polarizing beam splitters (PBS) and single-photon detectors (SPD).}\label{fig:experiment}
\end{figure}
\begin{figure*}[t!] 
\includegraphics[width=180mm]{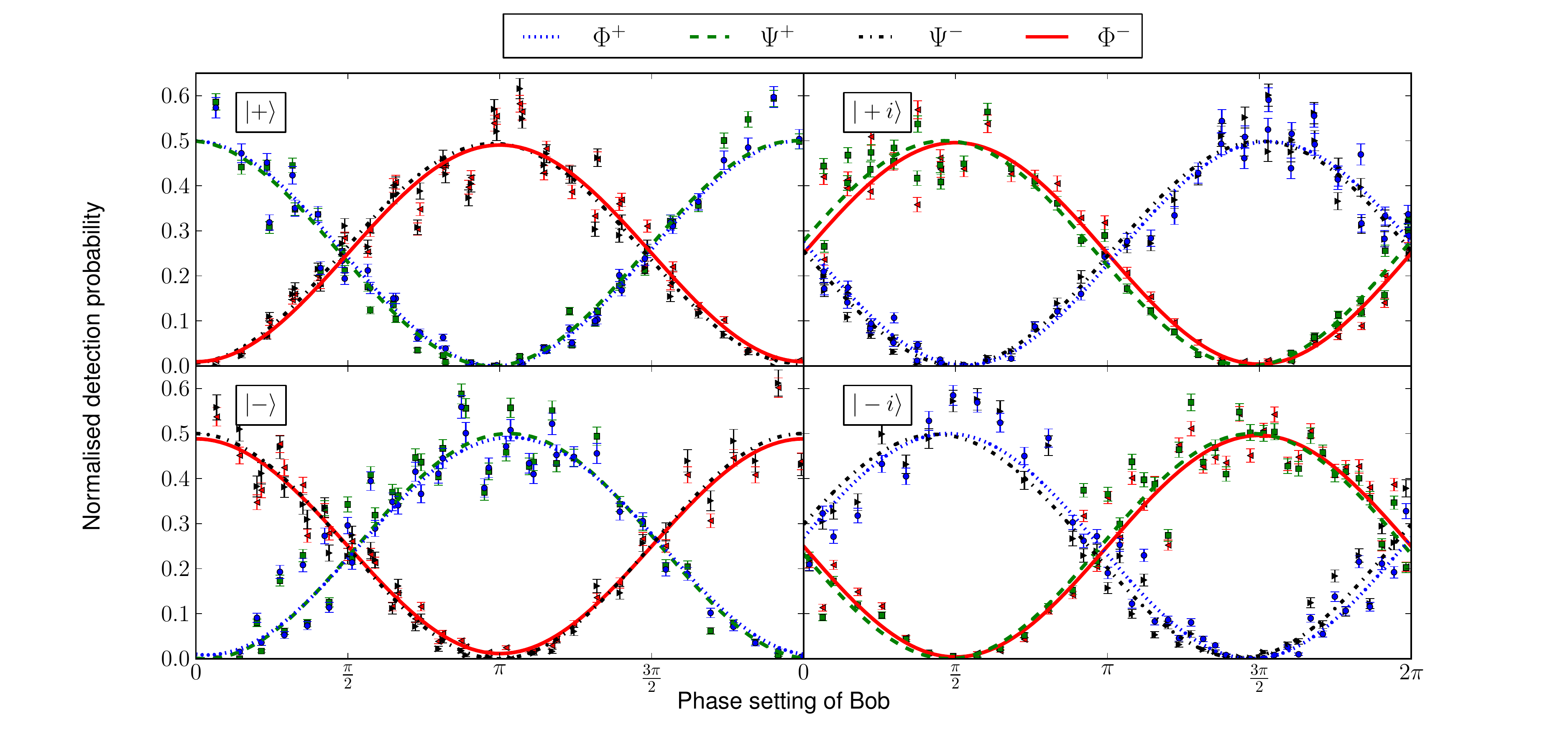}
\caption{Experimental Bell-state measurement outcomes as a function of the phase setting inside Bob's interferometer. Four sets of measurements are shown, one for each of the possible states sent by Alice.} \label{fig:data_curves}
\end{figure*}

Finally, the security of ddiQKD follows directly from that of the BB84 QKD scheme, with the additional benefit that detectors are excluded from the security analysis. In particular, the security can be obtained by using the entropic uncertainty relation proof technique~\cite{Tomamichel2012, Lim2014}: in the asymptotic limit, and under the approximation that the BB84 polarization states are prepared correctly, the secret key fraction is $\propto 1-2h(Q)$, where $h$ is the binary entropy function and $Q$ is the error rate of the sifted key. In fact, the finite-key security performance of ddiQKD is expected to be similar to the one of the single-photon BB84~\cite{Tomamichel2012} since only single-photon detections are required on Bob's side. Likewise, for a more practical implementation using the decoy-state method for WCS, we expect the security performance to be similar to the one in Ref.~\cite{Lim2014}.

We implemented a proof-of-principle experiment as illustrated in Fig.~\ref{fig:experiment}. We started with the generation of a pair of correlated photons by type-0 SPDC in a fiber-pigtailed periodically-poled lithium-niobate waveguide (PPLN-WG)~\cite{Tanzilli2012}. The waveguide was pumped with a continuous wave diode laser (Toptica DL100) at 780~nm and the signal and idler photons were deterministically separated by dense wavelength division multiplexers at 1563.9~nm (200~GHz) and 1556.6~nm (100~GHz), respectively. The idler photon was detected by a free-running InGaAs single-photon detector (ID Quantique ID220). The polarization of the heralded signal photon was set to $\ket{+}$ before passing through a Soleil-Babinet, which allowed us to rotate the state around the equator of the Bloch sphere and prepare Alice's single-photon state. Bob's  device consisted of a balanced interferometer, with a polarization controller in the upper arm acting as a HWP and a piezo phase modulator in the lower arm. The outputs of the BSM corresponding to $\ket{\Phi^{-}}$ and $\ket{\Psi^{-}}$ were delayed by 2.5~ns before being combined using two PBSs (see Fig.~\ref{fig:experiment}) with the other two outputs, which allowed the use of two detectors for all four outcomes. Bob's free-running InGaAs SPDs were cooled with a Stirling cooler to $-90^{\tn{o}}\tn{C}$ and had a dark count rate of less than 50~cps at 25\% efficiency~\cite{Korzh2014a}. The detection events were recorded by a time-to-digital converter (TDC). The $g^{(2)}(0)$ of the single photons at Alice was about $10^{-3}$ in a 1~ns coincidence window. Due to the extremely low dark count probability of the InGaAs detectors, the probability of having a double detection at Bob was $< 10^{-6}$.

To analyze the detection outcomes for all combinations of Alice and Bob's settings, we fixed the state prepared by Alice and scanned the phase of Bob's interferometer. Figure~\ref{fig:data_curves} shows four curves, one for each of the polarization states chosen by Alice, representing the normalized probability of each Bell-state being announced at any given phase setting in Bob's interferometer. The measurement points were fitted in order to calculate the visibility, with the highest average value obtained being $99.2\pm1.5\%$ for the $\ket{-i}$ input state at Bob and the lowest value of $96.0\pm2.1\%$ for the $\ket{-}$ state. \del{In a complete implementation of the QKD protocol Bob would randomly select four phase settings corresponding to the four BB84 states as outlined in section~\ref{sec:concept}.} Table~\ref{tab:correlation} shows the theoretical Bell-state announcement probability for every combination of Alice and Bob's settings. We complete this correlation table with the experimental results by selecting points from Fig.~\ref{fig:data_curves} closest to the desired settings for Bob. One can see that the experimental values coincide with the prediction and the overall quantum bit error rate, $Q$, was $1.5\pm0.5\%$. The total detection rate was around 60~cps.

While the concept of ddiQKD is fundamentally the same as mdiQKD, some subtleties need to be pointed out. For instance, in mdiQKD, Eve can interact with Alice's and Bob's qubits, but in our scheme only with Alice's qubit. Furthermore, we extend the trusted device boundary in Bob's laboratory to include the linear optical elements of the BSM, leaving only the single-photon detectors as untrusted devices. This means that Eve can have full control over their functionalities, e.g. she can control the response functions of the detectors~\cite{Liu2014}. But Bob can ensure that no additional information, other than the outcome of the BSM, leaks out of his lab. Indeed, if Eve had access to the output ports of Bob's PBSs she could carry out a Trojan-horse attack~\cite{Gisin2006} in order to gain information about the phase setting of Bob's interferometer. Note that attacks targeting the state preparation devices are also applicable to mdiQKD, but can be resolved (see Refs.~\cite{Gonzalez2014, Xu2014} for further discussion).

In practice, an implementation of ddiQKD using WCSs together with the decoy-state method could yield SKRs comparable with existing GHz clocked systems~\cite{Dixon2010, Tanaka2012, Walenta2014, Korzh2014c}. In particular, ultra-fast generation of polarization states could be achieved using a birefringence modulator scheme as used in Ref.~\cite{Lunghi2014}. We would like to highlight that the concept of TQSP entanglement employed in the ddiQKD scheme can be achieved by using any two DOFs of the single-photon. For example, Alice could encode a time-bin qubit~\cite{Korzh2013} followed by Bob's addition of a polarization qubit to the same photon.
 
In summary we implemented a ddiQKD protocol that overcomes the main disadvantages of the mdiQKD protocol whilst offering the same level of security. Future theoretical work should focus on deriving a bound on the extractable key length in a finite key scenario. This work paves the way to practical, high-performance and detector-side-channel free QKD. 

We would like to acknowledge Gustavo Lima, Guilherme Xavier and Marcos Curty for stimulating discussions regarding the basic idea. We thank ID Quantique and Battelle for the PPLN-WG and the Swiss NCCR QSIT for financial support.
%
\end{document}